\documentclass[prd,twocolumn,showpacs,showkeys,floatfix]{revtex4}
\usepackage{amsmath,amssymb}
\usepackage{latexsym}
\usepackage{bm}
\usepackage{hyperref}
\makeatletter
\def\eqnarray{\stepcounter{equation}\let\@currentlabel=\theequation
\global\@eqnswtrue
\global\@eqcnt\z@\tabskip\@centering\let\\=\@eqncr
$$\halign to \displaywidth\bgroup\@eqnsel\hskip\@centering
  $\displaystyle\tabskip\z@{##}$&\global\@eqcnt\@ne
  \hfil${\;##\;}$\hfil
  &\global\@eqcnt\tw@ $\displaystyle\tabskip\z@{##}$\hfil
   \tabskip\@centering&\llap{##}\tabskip\z@\cr}
\begin{document}
\title{DYNAMICAL INSTABILITY AND THE EXPANSION-FREE CONDITION}
\author{L. Herrera} 
\email{laherrera@cantv.net.ve. Also at U.C.V., Caracas}
\affiliation{Departamento   de F\'{\i}sica Te\'orica e Historia de la  Ciencia,
Universidad del Pa\'{\i}s Vasco, Bilbao, Spain}
\author{G. Le Denmat}
\email{gerard.le\_denmat@upmc.fr}
\affiliation{Observatoire de Paris, Universit\'e Pierre et Marie Curie,LERMA(ERGA) CNRS - UMR 8112, 94200 Ivry, France.}
\author{N.O. Santos}
\email{N.O.Santos@qmul.ac.uk}
\affiliation{School of Mathematical Sciences, Queen Mary
University of London, London E1 4NS, UK. and\\
 Observatoire de Paris, Universit\'e Pierre et Marie Curie,LERMA(ERGA) CNRS - UMR 8112, 94200 Ivry, France.}
\begin{abstract}
We study the dynamical instability of  a  spherically symmetric anisotropic fluid which collapses adiabatically under the condition of vanishing expansion scalar. The Newtonian and post Newtonian regimes are considered in detail. It is shown that  within those two approximations the adiabatic index $\Gamma_1$, measuring the fluid stiffness, does not  play any role. Instead,   the range of instability is determined by the anisotropy of the fluid pressures and the radial profile of  the energy density, independently of its stiffness, in a way which is fully consistent  with results previously obtained from the study on the Tolman mass.
\end{abstract}
\date{\today}
\pacs{04.40.-b, 04.20.-q, 04.40.Dg, 04.40.Nr}
\keywords{Relativistic fluids, stability, local anisotropy of pressure.}
\maketitle

\newpage

\section{Introduction}
The problem of stability  is of the utmost relevance in the study of Newtonian and general relativistic models of self-gravitating objects. This  becomes evident if we recall that any static stellar model, in order to be of any use, has to be stable against fluctuations. Furthermore, such a problem is closely related to the one of structure formation, since different degrees of stability/instability will lead to different patterns of evolution in the collapse of self-gravitating objects. It is therefore not surprising that a great deal of work has been devoted to this issue  since the pioneering paper by Chandrasekhar \cite{Ch}.

Extensions of Chandrasekhar's result  to  non-adiabatic fluids \cite{MN1,Chan}, anisotropic fluids \cite{Chan1} and shearing viscous fluids \cite{Chan2}, have been carried out in the past. In all of these works the key variable is the adiabatic index  $\Gamma_1$, whose  value  defines the range of instability.  Thus for a Newtonian perfect fluid, the system is unstable for $\Gamma_1<4/3$. In the above mentioned references it was shown how different physical aspects of the fluid affect the range of instability of the system. More recently \cite{Horvat} the stability of anisotropic stars  with quasi-local equation of state \cite{q1, q2} has been invetigated.

This work represents another forward step in that direction.  Our main goal here consists in studying the dynamical instability of a spherically symmetric fluid distribution, under the assumption of vanishing expansion scalar.

The main motivation to undertake such an endeavour  is provided by the following argument:
Highly energetic explosions in self-gravitating fluid distributions are  common events in relativistic astrophysics (see for example \cite{1N,2N} and references therein). Accordingly, a relevant question related to this issue is
\begin{itemize}
\item How the system evolves after the explosion?
\end{itemize}

Now, in a recent series of paper it has been stablished that the expansion-free condition is particularly suitable for describing   that kind of phenomena \cite{H1, H2, anali}.

Indeed, since the expansion scalar describes the rate of change  of   small volumes of the fluid, it  is intuitively clear that the evolution of an expansion--free  spherically symmetric fluid distribution is consistent with  the existence of a vacuum cavity  within the distribution which would be formed after a central explosion. This point is further discussed in this paper. Therefore potential applications of our results are expected for those astrophysical scenarios where a cavity within the fluid distribution is likely to be present (see for example \cite{cavity}). Furthermore, as the fluid (under the expansion--free condition) reaches the centre entails a blowup of the shear scalar, whose consequences in the appearance of a naked singularity cannot be overlooked \cite{s3p}.

The first known model satisfying the expansion-free condition is due to Skripkin \cite{Skripkin} for  the particular case of a non-dissipative isotropic fluid, with constant energy density. In that work, Skripkin addressed the very interesting problem of the evolution of a  spherically symmetric fluid  distribution  following a central explosion. As a result of the conditions imposed by him a Minkowskian  cavity should surround the centre of the fluid distribution.

Only recently this model has been again addressed. In \cite{H1} a general study on shearing with vanishing expansion scalar $\Theta$ of spherical fluid evolution is presented, which includes pressure anisotropy and dissipation.
While in \cite{H2} it is shown that the Skripkin model is incompatible with Darmois' junction conditions \cite{26}, and that inhomogeneous expansion-free dust  models are deprived of physical interest since they imply negative energy density distributions. Further  analytical solutions describing expansion--free evolution  may be found in \cite{anali}. Cavity evolution under kinematical conditions other than expansion-free, have been considered in \cite{Herrera}.

From  the above reasons, it follows that most, physically meaningful, expansion-free models should require anisotropy in the pressure and energy density inhomogeneity. Thus we shall  consider in this paper locally anisotropic fluids (further arguments to justify such kind of fluid distributions may be found in \cite{Herreraanis,Hetal,Ivanov} and references therein). Also, even though it is already an established fact, that gravitational collapse is a
highly dissipative process (see \cite{matter,Mitra} and references
therein), we shall restrain here for simplicity to adiabatic evolution. The role played by different dissipative processes in the instability of a fluid distribution have been discussed in detail in \cite{MN1,Chan,Chan1,Chan2}.

The fluid under consideration has two delimiting hypersurfaces. The external one separating the fluid distribution from a vacuum Schwarzschild spacetime  and the  internal one, delimiting the cavity within which there is Minkowski spacetime. Thus,  we have to consider junction conditions on both hypersurfaces.

It should be mentioned that for cavities with sizes of the order of 20 Mpc or smaller, the assumption of  a  spherically symmetric spacetime outside the cavity is quite reasonable, since the observed universe cannot be considered homogeneous on scales less than 150-300 Mpc. However for  larger cavities it should be more appropriate to consider their embedding in an expanding Lema\^{\i}tre-Friedmann-Robertson-Walker spacetime (for the specific case of void modeling in expanding universes see \cite{bill,Torres} and references therein).

Next we shall present our perturbative scheme which is very similar to that employed in \cite{MN1,Chan,Chan1,Chan2}, but now with the additional expansion-free condition. The study of  the resulting equations shows that at Newtonian and post Newtonian approximations, the range of instability is independent on $\Gamma_1$. This result is intuitively clear if we recall that, on the one hand the expansion-free condition implies, roughly speaking, that (at least close to Newtonian regime), the fluid evolves  without  being ``compressed'', and on the other  that  $\Gamma_1$  somehow measures the (in)compressibility  of the fluid. Therefore it is to be expected   that not very far from Newtonian regime $\Gamma_1$  does not appear in the discussion under expansion-free condition. Beyond post Newtonian approximation, intuition  is no longer a reliable guide and $\Gamma_1$ is expected  to play a role in the discussion.

The plan of the paper is as follows. In section II we give the energy-momentum tensor, the field equations and the junction conditions. Section III is devoted to explain the physical implications of the expansion-free evolution. The perturbation scheme is presented in section IV. In  section V a brief discussion on Newtonian and post Newtonian approximations is presented. The dynamical expansion-free equation is obtained in section VI. Finally there is a concluding section.

\section{The energy-momentum tensor, the field
 equations and junction conditions}
We consider a spherically symmetric distribution  of collapsing
fluid, bounded by a spherical surface  $\Sigma^{(e)}$.The fluid is
assumed to be locally anisotropic.
Choosing comoving coordinates inside  $\Sigma^{(e)}$, the general
interior metric can be written
\begin{equation}
ds^2_-=-A^2dt^2+B^2dr^2+R^2(d\theta^2+\sin^2\theta d\phi^2),
\label{1}
\end{equation}
where $A$, $B$ and $R$ are functions of $t$ and $r$ and are assumed
positive. We number the coordinates $x^0=t$, $x^1=r$, $x^2=\theta$
and $x^3=\phi$. Observe that $A$ and $B$ are dimensionless, whereas $R$ has the same dimension as $r$. Two radii are determined for a collapsing spherical fluid distribution by the metric (\ref{1}). The first is determined by $R(t,r)$ representing the radius as measured by its spherical surface, hence called its {\it areal radius}. The second is obtained out its radial integration 
$\int B(t,r)dr$, hence called {\it proper radius}. These two radii, in Einstein's theory, in general need not to be equal  unlike in Euclidean  geometry (fluids with both radii equal are studied in \cite{Herrera1}).

The matter energy-momentum $T_{\alpha\beta}^-$ inside $\Sigma^{(e)}$
has the form
\begin{equation}
T_{\alpha\beta}^-=(\mu +P_{\perp})V_{\alpha}V_{\beta}+P_{\perp}g_{\alpha\beta}+(P_r-P_{\perp})\chi_{\alpha}\chi_{\beta}, \label{2}
\end{equation}
where $\mu$ is the energy density, $P_r$ the radial pressure, $P_{\perp}$ the tangential pressure,
$V^{\alpha}$ the four-velocity of the fluid and $\chi_{\alpha}$ a unit four-vector along the radial direction.
These quantities satisfy
\begin{equation}
V^{\alpha}V_{\alpha}=-1\;\;, \;\; \chi^{\alpha}\chi_{\alpha}=1\;\;, \;\;  \chi^{\alpha}V_{\alpha}=0. \label{3}
\end{equation}
The four-acceleration $a_{\alpha}$ and the expansion $\Theta$ of the fluid are
given by
\begin{equation}
a_{\alpha}=V_{\alpha ;\beta}V^{\beta}, \;\;
\Theta={V^{\alpha}}_{;\alpha}, \label{4}
\end{equation}
and its  shear $\sigma_{\alpha\beta}$ by
\begin{equation}
\sigma_{\alpha\beta}=V_{(\alpha
;\beta)}+a_{(\alpha}V_{\beta)}-\frac{1}{3}\Theta(g_{\alpha\beta}+V_{\alpha}V
_{\beta}).
\label{5}
\end{equation}
Since we assumed the metric (\ref{1}) comoving then
\begin{equation}
V^{\alpha}=A^{-1}\delta_0^{\alpha}, \;\;
\;\; \chi^{\alpha}=B^{-1}\delta^{\alpha}_1. \label{6}
\end{equation}

From  (\ref{4}) with (\ref{6}) we have the non zero component for the  four-acceleration and its scalar,
\begin{equation}
a_1=\frac{A^{\prime}}{A}, \;\; a^{\alpha}a_{\alpha}=\left(\frac{A^{\prime}}{AB}\right)^2, \label{7}
\end{equation}
and for the expansion
\begin{equation}
\Theta=\frac{1}{A}\left(\frac{\dot{B}}{B}+2\frac{\dot{R}}{R}\right),
\label{8}
\end{equation}
where the  prime stands for $r$
differentiation and the dot stands for differentiation with respect to $t$.
With (\ref{6}) we obtain
for the shear (\ref{5}) its non zero components
\begin{equation}
\sigma_{11}=\frac{2}{3}B^2\sigma, \;\;
\sigma_{22}=\sigma_{33}\sin^{-2}\theta=-\frac{1}{3}R^2\sigma,
 \label{10}
\end{equation}
and its scalar
\begin{equation}
\sigma^{\alpha\beta}\sigma_{\alpha\beta}=\frac{2}{3}\sigma^2,
\label{5bn}
\end{equation}
where
\begin{equation}
\sigma=\frac{1}{A}\left(\frac{\dot{B}}{B}-\frac{\dot{R}}{R}\right). \label{11}
\end{equation}

Einstein's field equations for the interior spacetime (\ref{1}) to $\Sigma^{(e)}$ are given by
\begin{equation}
G_{\alpha\beta}^-=\kappa T_{\alpha\beta}^-,
\label{12}
\end{equation}
 and its non zero components
with (\ref{1}), (\ref{2}) and (\ref{6})
become
\begin{eqnarray}
\kappa T_{00}^-=\kappa \mu A^2
=\left(2\frac{\dot{B}}{B}+\frac{\dot{R}}{R}\right)\frac{\dot{R}}{R}\nonumber\\
-\left(\frac{A}{B}\right)^2\left[2\frac{R^{\prime\prime}}{R}+\left(\frac{R^{\prime}}{R}\right)^2
-2\frac{B^{\prime}}{B}\frac{R^{\prime}}{R}-\left(\frac{B}{R}\right)^2\right],
\label{13} \\
\kappa T_{01}^-=0
=-2\left(\frac{{\dot R}^{\prime}}{R}
-\frac{\dot B}{B}\frac{R^{\prime}}{R}-\frac{\dot
R}{R}\frac{A^{\prime}}{A}\right),
\label{14} \\
\kappa T_{11}^-=\kappa P_rB^2 \nonumber\\
=-\left(\frac{B}{A}\right)^2\left[2\frac{\ddot{R}}{R}-\left(2\frac{\dot A}{A}-\frac{\dot{R}}{R}\right)
\frac{\dot R}{R}\right]\nonumber\\
+\left(2\frac{A^{\prime}}{A}+\frac{R^{\prime}}{R}\right)\frac{R^{\prime}}{R}-\left(\frac{B}{R}\right)^2,
\label{15} \\
\kappa T_{22}^-=\kappa T_{33}^-\sin^{-2}\theta=\kappa P_{\perp}R^2\nonumber \\
=-\left(\frac{R}{A}\right)^2\left[\frac{\ddot{B}}{B}+\frac{\ddot{R}}{R}
-\frac{\dot{A}}{A}\left(\frac{\dot{B}}{B}+\frac{\dot{R}}{R}\right)
+\frac{\dot{B}}{B}\frac{\dot{R}}{R}\right]\nonumber\\
+\left(\frac{R}{B}\right)^2\left[\frac{A^{\prime\prime}}{A}
+\frac{R^{\prime\prime}}{R}-\frac{A^{\prime}}{A}\left(\frac{B^{\prime}}{B}-\frac{R^{\prime}}{R}\right)
-\frac{B^{\prime}}{B}\frac{R^{\prime}}{R}\right].\label{16}
\end{eqnarray}
The component (\ref{14}) can be rewritten with (\ref{8}) and
(\ref{11}) as
\begin{equation}
\frac{1}{3}(\Theta-\sigma)^{\prime}
-\sigma\frac{R^{\prime}}{R}=0.\label{17}
\end{equation}

The mass function $m(t,r)$ introduced by Misner and Sharp
\cite{Misner} (see also \cite{Cahill}) reads
\begin{equation}
m=\frac{R^3}{2}{R_{23}}^{23}
=\frac{R}{2}\left[\left(\frac{\dot R}{A}\right)^2-\left(\frac{R^{\prime}}{B}\right)^2+1\right],
\label{18}
\end{equation}
and the non trivial components of the Bianchi identities, $T_{;\beta}^{-\alpha\beta}=0$, from (\ref{12}) yield
\begin{equation}
T_{;\beta}^{-\alpha\beta}V_{\alpha}=-\frac{1}{A}\left[{\dot\mu}+(\mu +P_r)\frac{\dot B}{B}
+2(\mu+P_{\perp})\frac{\dot R}{R}\right] =0, \label{17a}
\end{equation}
or using (\ref{8})
\begin{eqnarray}
{\dot\mu}+(\mu +P_r)A\Theta
+2(P_{\perp}-P_r)\frac{\dot R}{R}=0,
\label{17abis}
\end{eqnarray}
and
\begin{equation}
T_{;\beta}^{-\alpha\beta}\chi_{\alpha}=
\frac{1}{B}\left[P^{\prime}_r
+(\mu +P_r)\frac{A^{\prime}}{A}+2(P_r-P_{\perp})\frac{R^{\prime}}{R}\right]=0. \label{17b}
\end{equation}

Introducing  the proper time derivative $D_T$
given by
\begin{equation}
D_T=\frac{1}{A}\frac{\partial}{\partial t}, \label{16N}
\end{equation}
 we can define the velocity $U$ of the collapsing
fluid (for another definition of velocity see  section III) as the variation of the areal radius with respect to proper time, i.e.,
\begin{equation}
U=D_TR. \label{19a}
\end{equation}
Using field equations and (\ref{18}) we may write
\begin{eqnarray}
m^{\prime}=\frac{\kappa}{2}\mu R^{\prime}R^2.
\label{27Dr}
\end{eqnarray}

Outside $\Sigma^{(e)}$  we assume we have the Schwarzschild
spacetime, i.e.
\begin{equation}
ds^2=-\left(1-\frac{2M}{\rho}\right)dv^2-2d\rho dv+\rho^2(d\theta^2
+\sin^2\theta
d\phi^2) \label{19},
\end{equation}
where $M$  denotes the total mass,
and  $v$ is the retarded time.

The matching of the  adiabatic fluid  sphere  to
Schwarzschild
spacetime, on the surface $r=r_{\Sigma^{(e)}}=$ constant (or $\rho=\rho(v)_{\Sigma^{(e)}}$ in the coordinates  of (\ref{19})) , requires  the continuity of the first and second differential forms (Darmois conditions), implying
\begin{equation}
 Adt\stackrel{\Sigma^{(e)}}{=}dv \left(1-\frac{2M}{\rho}\right), \label{junction1fn}
\end{equation}
\begin{equation}
R\stackrel{\Sigma^{(e)}}{=}\rho(v), \label{junction1f2n}
\end{equation}
\begin{equation}
m(t,r)\stackrel{\Sigma^{(e)}}{=}M, \label{junction1}
\end{equation}
and
\begin{widetext}
\begin{eqnarray}
2\left(\frac{{\dot R}^{\prime}}{R}-\frac{\dot B}{B}\frac{R^{\prime}}{R}-\frac{\dot R}{R}\frac{A^{\prime}}{A}\right)
\stackrel{\Sigma^{(e)}}{=}-\frac{B}{A}\left[2\frac{\ddot R}{R}
-\left(2\frac{\dot A}{A}
-\frac{\dot R}{R}\right)\frac{\dot R}{R}\right]+\frac{A}{B}\left[\left(2\frac{A^{\prime}}{A}
+\frac{R^{\prime}}{R}\right)\frac{R^{\prime}}{R}-\left(\frac{B}{R}\right)^2\right],
\label{j2}
\end{eqnarray}
\end{widetext}
where $\stackrel{\Sigma^{(e)}}{=}$ means that both sides of the equation
are evaluated on $\Sigma^{(e)}$.
Comparing (\ref{j2}) with  (\ref{13}) and (\ref{14}) one obtains
\begin{equation}
P_r\stackrel{\Sigma^{(e)}}{=}0.\label{j3}
\end{equation}
Thus   the matching of
(\ref{1})  and (\ref{19}) on $\Sigma^{(e)}$ produces (\ref{junction1}) and  (\ref{j3}).

As we mentioned in the introduction, the expansion-free models  present an internal vacuum cavity (reasons for the formation of this cavity are discussed in the following section.). If we call $\Sigma^{(i)} $ the boundary surface between the cavity and the fluid, then the matching of the Minkowski spacetime within the cavity to the fluid distribution, implies
\begin{equation}
m(t,r)\stackrel{\Sigma^{(i)}}{=}0, \label{junction1i}
\end{equation}
\begin{equation}
P_r\stackrel{\Sigma^{(i)}}{=}0.\label{j3i}
\end{equation}

\section{On the physical implications of the expansion-free evolution}
In section II we introduced the  variable $U$ which, as mentioned before, measures the variation of the areal radius $R$  per unit proper time.
Another possible definition of ``velocity'' may be introduced, as  the variation of the infinitesimal proper radial distance between two neighboring points ($\delta l$) per unit of proper time, i.e. $D_T(\delta l)$.
Then, it can be shown that (see \cite{H1} for details)
\begin{equation}
 \frac{D_T(\delta l)}{\delta l}= \frac{1}{3}(2\sigma +\Theta),
\label{vel15}
\end{equation}
or, by using (\ref{8}) and (\ref{11}),
\begin{equation}
 \frac{D_T(\delta l)}{\delta l}= \frac{\dot B}{AB}.
\label{vel16}
\end{equation}
Then with (\ref{8}), (\ref{11}), (\ref{19a}) and (\ref{vel16}) we can write
\begin{equation}
 \sigma= \frac{D_T(\delta l)}{\delta l}-\frac{D_T R}{R}= \frac{D_T(\delta l)}{\delta l}-\frac{U}{R},
\label{vel17}
\end{equation}
and
\begin{equation}
 \Theta= \frac{D_T(\delta l)}{\delta l}+\frac{2D_T R}{R}=\frac{D_T(\delta l)}{\delta l}+\frac{2U}{R},
\label{vel17bis}
\end{equation}
Thus the ``circumferential'' (or ``areal'') velocity $U$,  is  related to the change  of areal radius $R$  of a layer of matter, whereas $D_T(\delta l)$, has also the meaning of ``velocity'', being the relative velocity between neighboring layers of matter, and can be in general different from $U$.

We shall see now that the condition $\Theta=0$ is associated to  the existence of a cavity surrounding the centre of the fluid distribution.
Indeed, if $\Theta=0$ then  it follows from (\ref{vel17}) and (\ref{vel17bis}) that
\begin{equation}
\sigma=-3\frac{U}{R},\label{ns1}
\end{equation}
and feeding back (\ref{ns1}) into (\ref{17})  we get
\begin{equation}
\frac{U^{\prime}}{U}=-2\frac{ R^{\prime}}{R},
\label{27}
\end{equation}
whose integration with respect to $r$ yields
\begin{equation}
U=\frac{\zeta(t)}{R^2},
\label{28relvel}
\end{equation}
where $\zeta$ is an integration function of  $t$, implying
\begin{equation}
\sigma=-\frac{3\zeta(t)}{R^3}.
\label{28relveln}
\end{equation}

In the case when the fluid fills all the sphere, including the centre ($R(t,0)=0$), we have to impose the regularity condition $\zeta=0$, implying $U=0$.
Therefore if we want the expansion--free condition {\it applied to all fluid elements} to be compatible with a time dependent situation ($U\neq0$) we must assume that  either
\begin{itemize}
\item the fluid has no symmetry centre, or\\
\item the centre is surrounded by  a compact spherical section of another spacetime, suitably matched to the rest of the fluid.
\end{itemize}
Here we discard the first possibility since we are particularly interested in describing localized objects without the unusual topology of a spherical fluid without a centre. Also, within the context of the second alternative we have chosen an inner vacuum Minkowski spherical vacuole.

Thus the kinematical condition  $\Theta=0$ is consistent with an evolving spherically symmetric fluid  if there is a vacuum cavity surrounding its centre.

We can also arrive at this latter conclusion by the following qualitative argument. Since the expansion scalar describes the rate of change of small volumes of the fluid, it  is intuitively clear that in the case of an overall expansion (contraction), the increase (decrease) in  volume due to the increasing (decreasing)  area of  the external boundary surface must be compensated with  the increase (decrease) of the   area of the internal boundary surface (delimiting the cavity) in order to keep $\Theta$ vanishing.

\section{The perturbative scheme}
We shall now describe in some detail the perturbative scheme which will provide us with the main equation required for our (in)stability analysis.

We assume that initially the fluid is in static equilibrium, which means that the fluid is described by quantities only with radial coordinate dependence. These quantities are denoted by a subscript zero. We further suppose as usual, that the metric functions $A(t,r)$, $B(t,r)$ and $R(t,r)$ have the same time dependence in their perturbations. Therefore we consider the metric functions and material functions given by
\begin{eqnarray}
A(t,r)=A_0(r)+\epsilon T(t)a(r), \label{s1}\\
B(t,r)=B_0(r)+\epsilon T(t)b(r), \label{s2}\\
R(t,r)=R_0(r)+\epsilon T(t)c(r), \label{s3}\\
\mu (t,r)=\mu_0(r)+\epsilon {\bar\mu}(t,r), \label{s4}\\
P_r(t,r)=P_{r0}(r)+\epsilon {\bar P}_r(t,r), \label{s5}\\
P_{\perp}(t,r)=P_{\perp 0}(r)+\epsilon {\bar P}_{\perp}(t,r), \label{s6}\\
m(t,r)=m_0(r)+\epsilon {\bar m}(t,r), \label{s8}\\
\Theta(t,r)=\epsilon {\bar \Theta}(t,r), \label{s9}\\
\sigma(t,r)=\epsilon {\bar \sigma}(t,r), \label{s10}
\end{eqnarray}
where $0<\epsilon \ll 1$ and using the freedom allowed by the radial coordinate we choose the Schwarzschild coordinates with $R_0(r)=r$. Considering (\ref{s1}-\ref{s6}) we have from (\ref{13}-\ref{16})  for the static configuration
\begin{eqnarray}
\kappa\mu_0=\frac{1}{(B_0r)^2}\left(2r\frac{B_0^{\prime}}{B_0}+B_0^2-1\right), \label{s11}\\
\kappa P_{r0}=\frac{1}{(B_0r)^2}\left(2r\frac{A_0^{\prime}}{A_0}-B_0^2+1\right),\label{s12}\\
\kappa P_{\perp 0}=\frac{1}{B_0^2}\left[\frac{A_0^{\prime\prime}}{A_0}-\frac{A_0^{\prime}}{A_0}\frac{B_0^{\prime}}{B_0}
+\frac{1}{r}\left(\frac{A_0^{\prime}}{A_0}-\frac{B_0^{\prime}}{B_0}\right)\right]; \label{s13}
\end{eqnarray}
whereas from (\ref{13}-\ref{16})  we obtain for the perturbed quantities
\begin{widetext}
\begin{eqnarray}
\kappa{\bar\mu}=-2\frac{T}{B_0^2}\left[\left(\frac{c}{r}\right)^{\prime\prime}
-\frac{1}{r}\left(\frac{b}{B_0}\right)^{\prime}-
\left(\frac{B_0^{\prime}}{B_0}-\frac{3}{r}\right)\left(\frac{c}{r}\right)^{\prime}\right.
\left.-\left(\frac{B_0}{r}\right)^2
\left(\frac{b}{B_0}-\frac{c}{r}\right)\right]-2\kappa\mu_0T\frac{b}{B_0}, \label{s15}
\end{eqnarray}
\end{widetext}
\begin{eqnarray}
2\frac{{\dot T}}{A_0B_0}\left[\left(\frac{c}{r}\right)^{\prime}
-\frac{b}{rB_0}-\left(\frac{A_0^{\prime}}{A_0}-\frac{1}{r}\right)
\frac{c}{r}\right]=0, \label{s16}
\end{eqnarray}
\begin{widetext}
\begin{eqnarray}
\kappa{\bar P}_r=-2\frac{\ddot T}{A_0^2}\frac{c}{r}
+2\frac{T}{rB_0^2}\left[\left(\frac{a}{A_0}\right)^{\prime}
+\left(r\frac{A_0^{\prime}}{A_0}+1\right)\left(\frac{c}{r}\right)^{\prime}
-\frac{B_0^2}{r}\left(\frac{b}{B_0}-\frac{c}{r}\right)
\right]
-2\kappa P_{r0}T\frac{b}{B_0}, \label{s17}
\end{eqnarray}
\end{widetext}
\begin{widetext}
\begin{eqnarray}
\kappa {\bar P}_{\perp}=-\frac{\ddot T}{A_0^2}\left(\frac{b}{B_0}+\frac{c}{r}\right)
+\frac{T}{B_0^2}\left[\left(\frac{a}{A_0}\right)^{\prime\prime}
+\left(\frac{c}{r}\right)^{\prime\prime}+
\left(2\frac{A_0^{\prime}}{A_0}-\frac{B_0^{\prime}}{B_0}+\frac{1}{r}\right)
\left(\frac{a}{A_0}\right)^{\prime}\right. \nonumber\\
\left.-\left(\frac{A_0^{\prime}}{A_0}+\frac{1}{r}\right)\left(\frac{b}{B_0}\right)^{\prime}
+\left(\frac{A_0^{\prime}}{A_0}-\frac{B_0^{\prime}}{B_0}+\frac{2}{r}\right)
\left(\frac{c}{r}\right)^{\prime}\right]-2\kappa P_{\perp 0}T\frac{b}{B_0}; \label{s18}
\end{eqnarray}
\end{widetext}
and for the expansion (\ref{8}) and shear (\ref{11}) we have
\begin{equation}
{\bar\Theta}=\frac{\dot T}{A_0}\left(\frac{b}{B_0}+2\frac{c}{r}\right), \label{s20}
\end{equation}
\begin{equation}
{\bar\sigma}=\frac{\dot T}{A_0}\left(\frac{b}{B_0}-\frac{c}{r}\right). \label{s21}
\end{equation}

The Bianchi identities (\ref{17a}) and (\ref{17b}) become with (\ref{s1}-\ref{s6}), for the static configuration
\begin{eqnarray}
P^{\prime}_{r0}+(\mu_0+P_{r0})\frac{A_0^{\prime}}{A_0}+\frac{2}{r}(P_{r0}-P_{\perp 0})=0, \label{s22}
\end{eqnarray}
and for the perturbed configuration,
\begin{eqnarray}
\frac{1}{A_0}\left[{\dot{\bar\mu}}+(\mu_0+P_{r0}){\dot T}\frac{b}{B_0}+2(\mu_0+P_{\perp 0}){\dot T}
\frac{c}{r}\right] =0, \label{s23}
\end{eqnarray}
and
\begin{eqnarray}
\frac{1}{B_0}\left[{\bar P}^{\prime}_r+(\mu_0+P_{r0})T\left(\frac{a}{A_0}\right)^{\prime}
+({\bar\mu}+{\bar P}_r)\frac{A_0^{\prime}}{A_0}\right. \nonumber\\
\left.+2(P_{r0}-P_{\perp 0})T\left(\frac{c}{r}\right)^{\prime}
+2({\bar P}_r-{\bar P}_{\perp})\frac{1}{r}\right]=0. \label{s24}
\end{eqnarray}
By substituting (\ref{s16}) into (\ref{s23}) we can integrate it, and we find
\begin{equation}
{\bar\mu}=-\left[(\mu_0+P_{r0})\frac{b}{B_0}+2(\mu_0
+P_{\perp 0})\frac{c}{r}\right]T. \label{s24a}
\end{equation}

The total energy inside $\Sigma^{(e)}$ up to a radius $r$ given by (\ref{18}) with (\ref{s1}-\ref{s3}) and (\ref{s8}) becomes,
\begin{eqnarray}
m_0=\frac{r}{2}\left(1-\frac{1}{B_0^2}\right), \label{s25}\\
{\bar m}=-\frac{T}{B_0^2}\left[r\left(c^{\prime}-\frac{b}{B_0}\right)+(1-B_0^2)\frac{c}{2}\right]. \label{s26}
\end{eqnarray}

From the matching condition (\ref{j3}) with (\ref{s5})  we have,
\begin{equation}
P_{r0}\stackrel{\Sigma^{(e)}}{=}0, \;\; {\bar P}_r\stackrel{\Sigma^{(e)}}{=}0. \label{s27}
\end{equation}
For $c\neq 0$, which is the case that we want to study,
with (\ref{s16}), (\ref{s17}) and (\ref{s27}) we have
\begin{equation}
{\ddot T}-\alpha T\stackrel{\Sigma^{(e)}}{=}0, \label{s28}
\end{equation}
where
\begin{eqnarray}
\alpha=\left(\frac{A_0}{B_0}\right)^2\left[\left(\frac{a}{A_0}\right)^{\prime}+\left(r\frac{A_0^{\prime}}{A_0}+1\right)\left(\frac{c}{r}\right)^{\prime}\right. \nonumber\\
\left.
-\frac{B_0^2}{r}\left(\frac{b}{B_0}-\frac{c}{r}\right)\right]\frac{1}{c}, \label{s29}
\end{eqnarray}
Solutions of (\ref{s28}) include  functions which oscillate (corresponding to stable systems) and those which do not (unstable ones). Since we are interested here in establishing the range of instability,  we confine our attention to the non oscillating ones, i. e. we assume that $a(r)$, $b(r)$ and $c(r)$ are such that on $r_{\Sigma^{(e)}}$, $\alpha_{\Sigma^{(e)}}>0$. Then,
\begin{equation}
T(t)=-\exp\left({\sqrt{\alpha_{\Sigma^{(e)}}}}\;t \right ), \label{sa29}
\end{equation}
representing a system that starts collapsing at $t=-\infty$ when $T(-\infty)=0$ and the system is static, and goes collapsing, diminishing its areal radius, while $t$ increases.

Considering the second law of thermodynamics and the same arguments as given in \cite{MN1,Chan,Chan1}, we can express a relationship between ${\bar P}_r$ and ${\bar \mu}$ given by
\begin{equation}
{\bar P}_r=\Gamma_1 \frac{P_{r0}}{\mu_0+P_{r0}}{\bar\mu}, \label{s31}
\end{equation}
where $\Gamma_1$ is the adiabatic index  which measures the variation of  pressure  related to a given variation of density, thereby measuring the stiffness of the fluid.   We consider it constant throughout the fluid distribution or, at least, throughout the region that we want to study. We recall that $\Gamma_1$ coincides with the ratio of the specific heats for perfect Maxwell-Boltzman gas \cite{KW,BD, price}.
\\
\section{Newtonian and post Newtonian terms}
Before dwelling to obtain the dynamical expansion--free equation, and since we are using relativistic units, the following comments should be helpful in order  to identify  the terms belonging to the Newtonian (N), post Newtonian (pN) and post post Newtonian (ppN) approximations. These terms are considered for the instability conditions stemming from the dynamical equation in the N and pN approximations.

Thus, for N approximation we assume.
\begin{equation}
\mu_0\gg P_{r0}, \;\; \mu_0\gg P_{\perp 0}. \label{zz32}
\end{equation}

For the metric coefficients, given in c.g.s. units, expanded up to pN approximation  become
\begin{equation}
A_0=1-\frac{{\mathcal G}m_0}{{\mathcal C}^2r}, \;\;
B_0=1+\frac{{\mathcal G}m_0}{{\mathcal C}^2r}, \label{zz33}
\end{equation}
where $\mathcal G$ is the gravitational constant and $\mathcal C$ is the speed of light.

Next, from (\ref{s12}) with (\ref{s25}) we have (in relativistic units)
\begin{equation}
\frac{A_0^{\prime}}{A_0}=\frac{\kappa P_{r0}r^3+2m_0}{2r(r-2m_0)}, \label{z32}
\end{equation}
and substituting into (\ref{s22}) it yields
\begin{equation}
P_{r0}^{\prime}=-\left[\frac{\kappa P_{r0} r^3+2m_0}{2r(r-2m_0)}\right](\mu_0+P_{r0})+\frac{2}{r}(P_{\perp 0}
-P_{r0}). \label{z33}
\end{equation}
Writing (\ref{z33}) in c.g.s. units it becomes
\begin{equation}
P_{r0}^{\prime}=-\mathcal{G}\left[\frac{\mathcal{C}^{-2}\kappa P_{r0}r^3+2m_0}
{2r(r-2\mathcal{C}^{-2}\mathcal{G}m_0)}\right](\mu_0+\mathcal{C}^{-2}P_{r0})
+\frac{2}{r}(P_{\perp 0}-P_{r0}). \label{z34}
\end{equation}
Expanding (\ref{z34}) up to terms of $\mathcal{C}^{-4}$ order we obtain
\begin{widetext}
\begin{eqnarray}
P_{r0}^{\prime}=-\mathcal{G}\frac{\mu_0m_0}{r^2}+\frac{2}{r}(P_{\perp 0}-P_{r0})
-\frac{\mathcal{G}}{\mathcal{C}^2r^3}\left(2\mathcal{G}\mu_0m_0^2+P_{r0}m_0r
+\frac{\kappa}{2}\mu_0P_{r0}r^4\right) \nonumber\\
-\frac{\mathcal{G}}{\mathcal{C}^4r^4}\left(4\mathcal{G}^2\mu_0m_0^3+2\mathcal{G}P_{r0}m_0^2r
+\mathcal{G}\kappa \mu_0P_{r0}m_0r^4+\frac{\kappa}{2}P_{r0}^2r^5\right). \label{z36}
\end{eqnarray}
\end{widetext}
Hence from (\ref{z36}) we have the following terms for the different orders of approximations:
\begin{eqnarray}
\mbox{N order: terms of order} \;\mathcal{C}^{0} ; \label{z37}\\
\mbox{pN order: terms of order} \;\mathcal{C}^{-2}; \label{z38}\\
\mbox{ppN order: terms of order} \;\mathcal{C}^{-4}. \label{z39}
\end{eqnarray}

\section{The dynamical expansion-free equation}
With the equations so far obtained we can build the dynamical equation that we specialize to the expansion-free condition which is the aim of our study. The obtention of the dynamical equation is done via (\ref{s24}).

The expansion-free condition $\Theta=0$ implies from (\ref{s20})
\begin{equation}
\frac{b}{B_0}=-2\frac{c}{r}, \label{37}
\end{equation}
which together with the adiabatic condition (\ref{s16}) produces,
\begin{equation}
2\frac{\dot T}{r^3B_0}\left(\frac{r^2c}{A_0}\right)^{\prime}=0, \label{38}
\end{equation}
implying
\begin{equation}
c=k\frac{A_0}{r^2}, \label{38a}
\end{equation}
where $k$ is a constant.
With (\ref{37}) we have for (\ref{s24a})
\begin{equation}
{\bar\mu}=2(P_{r0}-P_{\perp 0})T\frac{c}{r}, \label{39}
\end{equation}
showing that the perturbed energy density of the system stems  from the static background anisotropy.
This fact  is easily  understood if we recall that under the expansion free condition, as it follows from (\ref{17abis}), changes in $\mu$ for any given fluid element,  depend exclusively on the pressure anisotropy.

On the other hand, with (\ref{s31}) and (\ref{39}) we have
\begin{equation}
\bar{P}_r=2\Gamma_1\frac{P_{r0}}{\mu_0+P_{r0}}(P_{r0}
-P_{\perp 0})T\frac{c}{r}. \label{40}
\end{equation}

Two further useful relations we can obtain. One from (\ref{s22}),
\begin{equation}
\frac{A_0^{\prime}}{A_0}=-\frac{1}{\mu_0+P_{r0}}\left[P_{r0}^{\prime}+\frac{2}{r}(P_{r0}-P_{\perp 0})\right], \label{43}
\end{equation}
and another from (\ref{s11}) and (\ref{s25})
\begin{equation}
\frac{B_0^{\prime}}{B_0}=\frac{\kappa\mu_0r^3-2m_0}{2r(r-2m_0)}. \label{43a}
\end{equation}

We want to study the instability conditions for the expansion-free fluid up to the pN approximation, since intermediate calculations are rather long we shall  put them in an  Appendix.

Considering the N approximation of the dynamical equation (\ref{x40}) by using (\ref{x43}) and (\ref{x47}-\ref{x49}) and that terms with $P_{r0}/\mu_0$ being of ppN order it reduces to
\begin{equation}
3\kappa\mu_0+\kappa |P_{r0}^{\prime}|r+2\left(\alpha_{\Sigma^{(e)}}-\frac{21}{r^2}\right)\frac{m_0}{r}
=4\kappa(5P_{r0}-2P_{\perp 0}), \label{48}
\end{equation}
or, using(\ref{27Dr}) and rearranging terms
\begin{widetext}
\begin{eqnarray}
\frac{\kappa}{18} |P_{r0}^{\prime}|r^4+\frac{\alpha_{\Sigma^{(e)}}}{9}m_0r^2=\frac{2\kappa}{9}(5P_{r0}-2P_{\perp 0})r^3+\frac{\kappa}{6}\left(7\int^{r}_{r_{\Sigma^{(i)}}}\mu_0r^2dr-\mu_0 r^3\right)\label{48bis}
\end{eqnarray}
\end{widetext}
where we assume $P_{r0}^{\prime}<0$.

For the pN approximation we have
\begin{widetext}
\begin{eqnarray}
3\kappa \mu_0+\kappa |P^{\prime}_{r0}|r+2\left(\alpha_{\Sigma^{(e)}}-\frac{21}{r^2}\right)\frac{m_0}{r}
+2\kappa|P^{\prime}_{r0}|m_0+\kappa \alpha_{\Sigma^{(e)}}P_{r0} r^2
-\kappa \mu_0\left(3\frac{m_0}{r}-2\alpha_{\Sigma^{(e)}}rm_0\right) \nonumber\\
+6\left(\alpha_{\Sigma^{(e)}}-\frac{5}{r^2}\right)\left(\frac{m_0}{r}\right)^2
=4\kappa(5P_{r0}-2P_{\perp 0})+2\kappa(P_{r0}-2P_{\perp 0})\frac{m_0}{r}. \label{49}
\end{eqnarray}
\end{widetext}
Let us first consider  the N approximation. The first observation to be made is that in the absence of a single parameter such as $\Gamma_1$ the assessment of the instability range depends in a rather complicated way on different structural properties of the fluid, such as pressure anisotropy and the radial profile of the energy density.

Indeed, for the onset of instability we need (\ref{48bis}) to be satisfied, and since the two terms at the left of (\ref{48bis}) are positive, then instabilities may develop only if so is the combination of the two terms at the right of (\ref{48bis}). For that to happen it will be sufficient that $P_{r0}>(2/5)P_{\perp 0}$ and that the last term in (\ref{48bis}) be positive. Let us explore this possibility in some detail.

Thus, let us consider  an energy density profile of the form $\mu_0=\beta r^n$ where $\beta$ is a positive constant, and $n$ is also a constant whose value ranges in the interval $-\infty <n<\infty$. In this case the last term in (\ref{48bis}) becomes (for $n\neq-3$)
\begin{equation}
\frac{\kappa \beta}{6(3+n)}r^{n+3}\left[4-n-7\left(\frac{r_{\Sigma^{(i)}}}{r}\right)^{3+n}\right],
\label{cn1}
\end{equation}
whereas for the $n=-3$ case we obtain
\begin{equation}
\frac{\kappa \beta}{6}\left[7\log{\left(\frac{r}{r_{\Sigma^{(i)}}}\right)}-1\right].
\label{cn1log}
\end{equation}
Then the following  possibilities arise:
\begin{enumerate}

\item $n\leq0$, $n\neq-3$.

In this case  (\ref{cn1}) will be positive if
\begin{equation}
r>r_{\Sigma^{(i)}}\left(\frac{7}{4-n}\right)^{1/(n+3)}.
\label{c1}
\end{equation}

Thus the maximal range of instability decreases from  its value for $|n|$ close to $3$, for which  we have  (see case 3 below)
\begin{equation}
r>r_{\Sigma^{(i)}} e^{1/7}\approx r_{\Sigma^{(i)}}1.15
\label{c11dif}
\end{equation}
to
\begin{equation}
r>r_{\Sigma^{(i)}}\left(\frac{7}{4}\right)^{1/3}\approx r_{\Sigma^{(i)}}1.20
\label{c11}
\end{equation}
corresponding to $n=0$  (the incompressible fluid). On the other hand as $|n|$ increases from $|n|=3$  the maximal region of instability will embrace the whole fluid, since
\begin{equation}
\lim_{n \rightarrow -\infty} \left(\frac{7}{4-n}\right)^{1/(3+n)}=1
\label{lim}
\end{equation}
In other words,  for $n<0$, the range of instability increases as the absolute value of $n$ increases.

\item $n\geq 0$.

In this case the positivity of the last term in (\ref{48bis}) is, again, assured in the region
\begin{equation}
r>r_{\Sigma^{(i)}}\left(\frac{7}{4-n}\right)^{1/(n+3)}.
\label{c11nuec}
\end{equation}
The range of instability decreases with $n$, vanishing  for $n\geq4$.

\item $n=-3$.

In this particular case,  the positivity of the the last term in (\ref{48bis}) is assured in the region
\begin{equation}
r>r_{\Sigma^{(i)}} e^{1/7}\approx r_{\Sigma^{(i)}}1.15
\label{c11otro}
\end{equation}
\end{enumerate}

Of course, if $P_{r0}<(2/5)P_{\perp 0}$ (for the same energy density profile), the maximal region of instability diminishes.

At pN order the situation is essentially the same, with  the relativistic effects at first order taken into account.

Observe   that  $\Gamma_1$ does not play any role in the N and pN orders. In other words at N order, no matter how stiff is the material, the system will be unstable as long as (\ref{48}) is satisfied. This is at variance  with the result in the non-vanishing expansion case  when it appears that the range of instability is defined by $\Gamma_1<(4/3)+$anisotropic term (see \cite{Chan1} for details). A similar remark applies for the pN order.

As mentioned before the fact that   $\Gamma_1$ does not enter into (\ref{48}) and (\ref{49}) becomes intelligible when we recall that the expansion-free collapse (close to the Newtonian regime) proceeds without ``compression'' of the fluid. Accordingly the stiffness of  matter is irrelevant for the onset of  instabilities, the later  being dependent on the local anisotropy of pressure and energy density inhomogeneity.

It should be observed that in the process of collapse (under the expansion-free condition) the energy density may increase or decrease depending on the difference between the velocity of the inner and the outer boundary surface. This diference in turn is determined  by  the local anisotropy of pressure and the radial profile of the energy density.

\section{Conclusions}
We have seen so far that in the study of dynamical instability, under the expansion-free condition, at, both, the Newtonian and post Newtonian regimes, the range of instability is defined by the local anisotropy of pressure and the energy density radial profile, but not by the adiabatic index  $\Gamma_1$. This implies that in the above mentioned regimes the stiffness of the fluid, measured by $\Gamma_1$, which generally plays the central role in the definition of the instability range, is irrelevant here. This fact strengthens  further  the relevance of local anisotropy of pressure and energy density inhomogeneity in the structure and evolution of self-gravitating objects.

It should be stressed that any possible model is further constrained  by physical requirements such as positivity of energy density, sound speed less than light (energy density greater than pressure),  and stability of local oscillations modes.

The association of a cavity with the expansion-free fluid distribution implies that the above presented study  describes the instability range of the cavity (keeping the expansion-free condition). We observe that the perturbation of the cavity itself produces a similar equation to $T(t)$ as in (\ref{s28}) and it imposes a further constraint to the system 
$\alpha_{\Sigma^{(e)}}=\alpha_{\Sigma^{(i)}}$.

It is worth mentioning that the role played   by the anisotropy and the energy density inhomogeneity, as it follows from the previous section, is fully consistent  with the results obtained from the study on the influence of those two factors on the active gravitational (Tolman) mass \cite{T}, presented in \cite{inh}-\cite{act}.

Indeed, for the Tolman mass $m_T$ \cite{T} interior to a sphere of radius $r$, up to pN approximation, the following expression may be obtained in the static or quasi-static case (see equation (47) in \cite{inh}, or equation (58) in \cite{split} or equation (50) in \cite{act} and the discussion therein)
\begin{widetext}
\begin{equation}
m_T=(m_T)_{\Sigma^{(e)}} \left(\frac{r}{r_{\Sigma^{(e)}}}\right)^3
+\kappa r^3 \int^{r_{\Sigma^{(e)}}}_r\left[
\left(P_{r0}-P_{\perp 0}\right)\frac{1}{r}
- \frac{1}{2r^4} \int^r_{r_{\Sigma^{(i)}}}\mu_0^{\prime}\tilde{r}^3d\tilde{r}\right] dr,
\label{emtebisbis}
\end{equation}
\end{widetext}
where (\ref{zz33}) has been used.

From the above  it is evident that  $\mu_0^{\prime}<0$ and $P_{r0}>P_{\perp 0}$ increase the Tolman mass. That result, together with the  fact that the Tolman mass may be interpreted as the active gravitational mass, provide full support to the conclusions on the instability range obtained in the previous section.

It is worth noticing that  if we keep the expansion-free condition all along the collapse, then as soon as the fluid approaches the centre there will be a blowup of the shear scalar, as implied by (\ref{28relveln}). But it is shown in \cite{s3p}  that sufficiently
strong shearing effects near the singularity delay the formation of the apparent horizon, implying the appearance of a  naked singularity. In other words the expansion-free  condition provides a simple scenario for  naked singularity formation. The obvious relevance of such an effect strengthens further the interest of the problem discussed here.

Finally, let us mention that an extensions  of these results for  $f(r)$ gravity theory, have been recently presented \cite{saff1}, \cite{saff}.

\section*{Appendix}
Inspecting (\ref{40}) we see that $\bar{P}_r$ is of ppN order, as well as $\bar{\mu}A_0^{\prime}/A_0$ by considering (\ref{39}) and (\ref{40}), hence (\ref{s24}) together with (\ref{sa29}) and (\ref{s18}) reduces to
\begin{widetext}
\begin{eqnarray}
\kappa(\mu_0+P_{r0})r\left(\frac{a}{A_0}\right)^{\prime}+2\kappa(P_{r0}
-P_{\perp 0})r\left(\frac{c}{r}\right)^{\prime}-8\kappa P_{\perp 0}\frac{c}{r}
-\frac{2}{B_0^2}\left[\left(\frac{a}{A_0}\right)^{\prime\prime}
+\left(\frac{c}{r}\right)^{\prime\prime}+\left(2\frac{A_0^{\prime}}{A_0}-\frac{B_0^{\prime}}{B_0}+
\frac{1}{r}\right)\left(\frac{a}{A_0}\right)^{\prime}\right.\nonumber\\
\left.+\left(3\frac{A_0^{\prime}}{A_0}
-\frac{B_0^{\prime}}{B_0}+\frac{4}{r}\right)\left(\frac{c}{r}\right)^{\prime}\right]-
2\frac{\alpha_{\Sigma^{(e)}}}{A_0^2}\frac{c}{r}=0.\label{x40}
\end{eqnarray}
\end{widetext}
From (\ref{s17}) with (\ref{s28}), (\ref{37}) and (\ref{38a}) we have
\begin{eqnarray}
\left(\frac{a}{A_0}\right)^{\prime}=-k\frac{A_0}{r^2}\left
[2\kappa P_{r0}B_0^2+\left(\frac{A_0^{\prime}}{A_0}\right)^2-\frac{2}{r}\frac{A_0^{\prime}}{A_0}\right.
\nonumber\\
\left.+\frac{3}{r^2}(B_0^2-1)-\alpha_{\Sigma^{(e)}}\left(\frac{B_0}{A_0}\right)^2\right]. \label{x42}
\end{eqnarray}
The first three terms of (\ref{x40}) with (\ref{38a}) and (\ref{x42}) up to pN order become
\begin{widetext}
\begin{eqnarray}
\kappa (\mu_0+P_{r0})r\left(\frac{a}{A_0}\right)^{\prime}+2\kappa (P_{r0}-P_{\perp 0})
r\left(\frac{c}{r}\right)^{\prime}-8\kappa P_{\perp 0}\frac{c}{r}
=-\frac{\kappa k A_0}{r^2}\left\{2\kappa\mu_0 P_{r0}rB_0^2+2P_{r0}^{\prime}
+\frac{2}{r}(5P_{r0}-P_{\perp})\right.\nonumber\\
\left.+(\mu_0+P_{r0})r\left[\frac{3}{r^2}(B_0^2-1)-\alpha_{\Sigma^{(e)}}\left(\frac{B_0}{A_0}\right)^2\right]
\right\}
=-\frac{\kappa k}{r^2}\left\{2\kappa\mu_0P_{r0}r+\left(1-\frac{m_0}{r}\right)\left[2P_{r0}^{\prime}
+\frac{2}{r}(5P_{r0}-P_{\perp 0})\right]\right. \nonumber\\
\left.+(\mu_0+P_{r0})r\left[6\frac{m_0}{r^3}
-\alpha_{\Sigma^{(e)}}\left(1+3\frac{m_0}{r}\right)\right]+\mu_0r\left(\frac{6}{r^2}
-\frac{15}{2}\alpha_{\Sigma^{(e)}}\right)\left(\frac{m_0}{r}\right)^2\right\}.\label{x43}
\end{eqnarray}
\end{widetext}
Now we calculate the following terms,
\begin{widetext}
\begin{eqnarray}
\left(\frac{a}{A_0}\right)^{\prime\prime}+\left(2\frac{A_0^{\prime}}{A_0}-\frac{B_0^{\prime}}{B_0}
+\frac{1}{r}\right)\left(\frac{a}{A_0}\right)^{\prime}
=k\frac{A_0}{r^2}\left[2\kappa P_{r0}B_0^2\left(\frac{1}{r}-\frac{B_0^{\prime}}{B_0}\right)\right.
\left.-2\kappa P_{r0}^{\prime}B_0^2+\frac{2}{r}\left(\frac{A_0^{\prime}}{A_0}\right)^{\prime}-
\frac{2}{r}\frac{A_0^{\prime}}{A_0}\frac{B_0^{\prime}}{B_0}
\right. \nonumber \\ \left.+\frac{1}{r^2}\frac{A_0^{\prime}}{A_0}(5-9B_0^2)\right.
\left.-\frac{3}{r^2}\frac{B_0^{\prime}}{B_0}(B_0^2+1)+
\frac{9}{r^3}(B_0^2-1)\right]
+\alpha_{\Sigma^{(e)}}k\frac{A_0}{r^2}\left(\frac{B_0}{A_0}\right)^2\left(\frac{A_0^{\prime}}{A_0}
+\frac{B_0^{\prime}}{B_0}-\frac{1}{r}\right), \label{x44}
\end{eqnarray}
\end{widetext}
and
\begin{widetext}
\begin{eqnarray}
\left(\frac{c}{r}\right)^{\prime\prime}+\left(3\frac{A_0^{\prime}}{A_0}-\frac{B_0^{\prime}}{B_0}+
\frac{4}{r}\right)\left(\frac{c}{r}\right)^{\prime}
=k\frac{A_0}{r^3}
\left[\left(\frac{A_0^{\prime}}{A_0}\right)^{\prime}-\frac{A_0^{\prime}}{A_0}\frac{B_0^{\prime}}{B_0}
-\frac{11}{r}\frac{A_0^{\prime}}{A_0}+\frac{3}{r}\frac{B_0^{\prime}}{B_0}\right],\label{x45}
\end{eqnarray}
\end{widetext}
where (\ref{38a}) and (\ref{x42}) have been used. With (\ref{x44}) and (\ref{x45}) we can build
\begin{widetext}
\begin{eqnarray}
-\frac{2}{B_0^2}\left[\left(\frac{a}{A_0}\right)^{\prime\prime}
+\left(\frac{c}{r}\right)^{\prime\prime}+\left(2\frac{A_0^{\prime}}{A_0}-\frac{B_0^{\prime}}{B_0}+
\frac{1}{r}\right)\left(\frac{a}{A_0}\right)^{\prime}\right.
\left.+\left(3\frac{A_0^{\prime}}{A_0}
-\frac{B_0^{\prime}}{B_0}+\frac{4}{r}\right)\left(\frac{c}{r}\right)^{\prime}\right]-
2\frac{\alpha_{\Sigma^{(e)}}}{A_0^2}\frac{c}{r}\nonumber\\
=4\kappa k \frac{A_0}{r^2}\left[P_{r0}^{\prime}+P_{r0}\left(\frac{B_0^{\prime}}{B_0}
-\frac{1}{r}\right)\right]-6k\frac{A_0}{B_0^2r^3}\left[
\left(\frac{A_0^{\prime}}{A_0}\right)^{\prime}
-\frac{A_0^{\prime}}{A_0}\frac{B_0^{\prime}}{B_0}\right.\nonumber \\
\left.-(3B_0^2+2)\frac{1}{r}\frac{A_0^{\prime}}{A_0}-\frac{1}{r}B_0B_0^{\prime}+
\frac{3}{r^2}(B_0^2-1)\right]
-2\frac{\alpha_{\Sigma^{(e)}}k}{A_0r^2}\left(\frac{A_0^{\prime}}{A_0}+\frac{B_0^{\prime}}{B_0}\right).
\label{x46}
\end{eqnarray}
\end{widetext}
By using (\ref{zz33}), (\ref{43}) and (\ref{43a}) we obtain the following expressions up to pN order,
\begin{widetext}
\begin{eqnarray}
4\kappa k \frac{A_0}{r^2}\left[P_{r0}^{\prime}+P_{r0}\left(\frac{B_0^{\prime}}{B_0}
-\frac{1}{r}\right)\right]
=\frac{4\kappa k}{r^2}\left(P^{\prime}_{r0}-\frac{P_{r0}}{r}+\frac{\kappa}{2}r\mu_0P_{r0}-
\frac{m_0}{r}P^{\prime}_{r0}\right), \label{x47}
\end{eqnarray}
\end{widetext}
\begin{widetext}
\begin{eqnarray}
-6k\frac{A_0}{B_0^2r^3}\left[
\left(\frac{A_0^{\prime}}{A_0}\right)^{\prime}
-\frac{A_0^{\prime}}{A_0}\frac{B_0^{\prime}}{B_0}
-(3B_0^2+2)\frac{1}{r}\frac{A_0^{\prime}}{A_0}-\frac{1}{r}B_0B_0^{\prime}+
\frac{3}{r^2}(B_0^2-1)\right]
=-6\frac{k}{r^3}\left\{\frac{\mu_0^{\prime}}{\mu_0^2}\left[P^{\prime}_{r0}+\frac{2}{r}(P_{r0}-
P_{\perp 0})\right]\right. \nonumber\\
\left.-\left(1-3\frac{m_0}{r}\right)\frac{1}{\mu_0}\left[P_{r0}^{\prime\prime}-
\frac{2}{r^2}(P_{r0}-P_{\perp 0})+\frac{2}{r}(P_{r0}^{\prime}-P_{\perp 0}^{\prime})\right]
\right.
\left.+\frac{\kappa r^3\mu_0-2m_0}{2r^2\mu_0}\left[P^{\prime}_{r0}+\frac{2}{r}(P_{r0}-P_{\perp 0})\right]\right. \nonumber\\
\left.+\left(5-9\frac{m_0}{r}\right)\frac{1}{r\mu_0}\left[P_{r0}^{\prime}
+\frac{2}{r}(P_{r0}-P_{\perp 0})\right]\right.
\left.-\frac{\kappa \mu_0}{2}\left[1+\frac{m_0}{r}+\frac{3}{2}\left(\frac{m_0}{r}\right)^2\right]+
\frac{1}{r^2}\left[7\frac{m_0}{r}-5\left(\frac{m_0}{r}\right)^2\right]\right\}, \label{x48}
\end{eqnarray}
\end{widetext}
\begin{widetext}
\begin{eqnarray}
-2\frac{\alpha_{\Sigma^{(e)}}k}{A_0r^2}\left(\frac{A_0^{\prime}}{A_0}+\frac{B_0^{\prime}}{B_0}\right)
=-2\frac{\alpha_{\Sigma^{(e)}}k}{r^2}\left\{-\left(1+\frac{m_0}{r}\right)\frac{1}{\mu_0}
\left[P_{r0}^{\prime}+\frac{2}{r}(P_{r0}-P_{\perp 0})\right]\right. \nonumber\\
\left.+\frac{\kappa}{2}\left[1+3\frac{m_0}{r}+\frac{15}{2}\left(\frac{m_0}{r}\right)^2\right]r\mu_0-
\frac{1}{r}\left[\frac{m_0}{r}+3\left(\frac{m_0}{r}\right)^2\right]\right\} ,\label{x49}
\end{eqnarray}
\end{widetext}
some terms of ppN order appearing in some of the equations above will be excluded in the analysis of section VI.

\end{document}